# Graphene nanoplatelets induced tailoring in photocatalytic activity and antibacterial characteristics of MgO/graphene nanoplatelets nanocomposites

Aqsa Arshad,[1,2,a)] Javed Iqbal,[3,a)] M. Siddiq,[4] Qaisar Mansoor,[5] M. Ismail,[5] Faisal Mehmood,[1] M. Ajmal,[6] and Zubia Abid[1]

[1]*Department of Physics, International Islamic University, Islamabad, Pakistan*
[2]*Department of Physics, Durham University, Durham DH1 3LE, United Kingdom*
[3]*Laboratory of Nanoscience and Technology, Department of Physics, Quaid i Azam University, Islamabad, Pakistan*
[4]*Department of Chemistry, Quaid i Azam University, Islamabad, Pakistan*
[5]*Institute of Biomedical and Genetic Engineering (IBGE), Islamabad, Pakistan*
[6]*Institute of Chemical Sciences, Bahauddin Zakariya University, Multan 60800, Pakistan*

The synthesis, physical, photocatalytic, and antibacterial properties of MgO and graphene nanoplatelets (GNPs) nanocomposites are reported. The crystallinity, phase, morphology, chemical bonding, and vibrational modes of prepared nanomaterials are studied. The conductive nature of GNPs is tailored via photocatalysis and enhanced antibacterial activity. It is interestingly observed that the MgO/GNPs nanocomposites with optimized GNPs content show a significant photocatalytic activity (97.23% degradation) as compared to bare MgO (43%) which makes it the potential photocatalyst for purification of industrial waste water. In addition, the effect of increased amount of GNPs on antibacterial performance of nanocomposites against pathogenic micro-organisms is researched, suggesting them toxic. MgO/GNPs 25% nanocomposite may have potential applications in waste water treatment and nanomedicine due its multifunctionality.

## I. INTRODUCTION

Multifunctional nanomaterials are of extreme importance in the modern era of industrialization. Due to rapid industrial growth, there is constant confrontation of contaminated natural resources like water. At the same time health risks posed by pathogenic bacteria need to be controlled by novel methods other than traditional antibiotics. Material scientists are in continuous effort to present solutions to health hazards created by synthetic dyes and bacterial contaminations. Recently, metal oxides (with or without chemical modification) are vastly researched for the photocatalytic applications, i.e., to address water splitting and dye contaminated water remediation.[1–5] Magnesium oxide (MgO) is a wide band gap insulating metal oxide with bandgap energy 5–6 eV. It has been the focus of research both from theoretical and experimental point of view since decades.[6–10] Different nanoscale morphology based applications of MgO make it an important ceramic material. It has been used as a nanothermometer (Ga filled MgO nanotubes),[11] antibacterial agent,[12–14] substrate for high $T_c$ superconducting materials (HTSC), passive layer for high mobility transistors[15] and an excellent dielectric material.[16] The bandgap energy of nanoscale MgO, being an insulator is high, i.e., 5 eV which drags the attention towards making it optically active for applications like photo catalysis. Its surface modification is highly desirable to make its efficient use in adsorption of dyes, photo-oxidation catalysis (for waste water cleaning) and solar cells.[17–19] Interestingly, the same material MgO, being biosafe for healthy human cells and toxic for bacteria[13,20–22] provides a platform to further investigate its toxicity to pathogenic bacteria. The bare MgO shows negligibly low photocatalytic activity. The photocatalytic efficiency of MgO is restricted by its large bandgap energy and the quick recombination of charge carriers. This problem can be addressed by introducing an electron acceptor material with MgO. In this regard, carbon nanostructures are an attractive choice. To tailor above mentioned features and applications in a single material, we have combined MgO with graphene nanoplatelets (GNPs) (i.e., an insulating and a conducting phase) in this work.

Graphene nanoplatelets (GNPs) are new class of derivatives of graphene family with quasi two dimensional, $sp^2$ hybridized carbon atomic layers stacked upon each other. The high mechanical strengths, excellent thermal and electrical conductivities and unique energy dispersion relation near the Dirac points make graphene family materials a novel generation of carbon nanostructures. These features have opened new horizons for researchers to assemble graphene family materials with various organic and inorganic nanoscale units to tailor the combined properties in a novel fashion with target to achieve modern technological requirements in all the fields of science and technology.[23–34]

The above mentioned superior characteristics lead researchers to obvious use of GNPs with ceramics to incorporate their features in the form of nanocomposites. MgO, $Mg(OH)_2$, and graphene family nanocomposites have been of very recent interest for researchers. There have been recent reports depicting nanocomposites as heat transfer

---
[a)]Authors to whom correspondence should be addressed. Electronic addresses: aqsa.arshad@iiu.edu.pk and javed.saggu@qau.edu.pk

efficient materials and adsorbents of organic dyes, etc.[35–37] These few reports have covered only the limited physical and chemical aspects of MgO and graphene family nanocomposites. There is much potential in the area for the future work. There is the need to ponder over various aspects of MgO/GNPs nanocomposites to further throw light on those properties which have not been researched yet. This work is focussed to achieve a high photocatalytic activity of MgO/GNPs nanocomposite as compared to MgO. In parallel, this study presents a comprehensive analysis of antibacterial efficiency of MgO/GNPs nanocomposites. To the best of our knowledge, here we report for the first time, the impact of MgO/GNPs nanocomposites on photodegradation of methyl orange. Moreover, the present report is the first article on GNPs loading dependent antibacterial properties of MgO/GNPs nanocomposites.

## II. EXPERIMENTAL METHOD

### A. Synthesis of nanocomposites

Here a versatile, low cost, wet chemical and surfactant free route for the preparation of MgO and MgO/GNPs nanocomposites is reported. All the chemicals were of analytical grade. Magnesium nitrate hexahydrate (99%, Merck), sodium hydroxide (>98%, Merck), ethyl alcohol (99%, Merck), GNPs (100%, KNano), and distilled water were used to synthesize the MgO and its nanocomposites.

The nanocomposites were synthesized by sonication assisted solvothermal method. The GNPs were dispersed in a mixed solvent of absolute ethanol and distilled water (1:1) by sonication for few hours at room temperature. At this stage 14.74 g of $Mg(NO_3)_2 \cdot 6H_2O$ was dispersed in the above solution, followed by sonication. Basic solution of NaOH was prepared in a mixed solvent of distilled water and absolute ethanol (1:1). The sonicated solution was added to the basic solution in a controlled manner, followed by vigorous magnetic stirring at 1200 rpm. This solution was transferred to tightly sealed Tefl lined autoclave. The autoclave was transferred to a pre-heated electric oven at 180 C for 10 h. After cooling down the autoclave naturally at room temperature, the collected material was washed several times using distilled water and ethanol. The precipitates so obtained were dried in an electric oven at 100ºC for 2 h. To completely transform the $Mg(OH)_2$ phase to MgO cubic phase in the composite material, it was given a post annealing session at 500ºC for 3 h in a tube furnace. The MgO nanohexagons were prepared by following same route but without the addition of GNPs. Two nanocomposites labelled as MgO/GNPs 12% and MgO/GNPs 25% were prepared with different loadings of GNPs, i.e., 12% and 25% of MgO, respectively.

### B. Photocatalytic activity experiment

The photocatalytic performance of the prepared samples was studied by monitoring the photo degradation of methyl orange in a photocatalytic chamber equipped with a 90 W, type-C UV lamp, as the irradiation source. In each experiment, 40 mg of photocatalyst was dispersed in a 100 ml of $1.5 \times 10^{-5}$ M methyl orange aqueous solution. The solution was magnetically stirred in dark for 30 min to achieve the adsorption-desorption equilibrium. Thereafter, the solution was exposed to the UV light and 4 ml of the exposed solutions was collected at regular intervals of 30 min. The collected samples were centrifuged to remove the photocatalysts. Then the samples were analyzed for the photocatalytic degradation using a UV-Vis spectrophotometer (Shimadzu Pharmaspec-1700).

### C. Preparation method for antibacterial activity

To investigate the antibacterial test, the method followed is same as reported previously.[38] Briefly the test samples (MgO/GNPs 12% and MgO/GNPs 25%) with a concentration of 10 mg/ml of sterile water were mixed and sonicated. The 200 μl of the test sample was added to 5 ml Luria- Bertani (LB) medium. 100μl of the inoculum (bacterial cul- ture in LB) was added to the above modified growth medium. The inoculated media containing the test sample was incubated at 37ºC for 24 h.

### D. Characterizations

The prepared materials were characterized for their various physio-chemical properties. The X-ray diffractograms were recorded using the Panalytical X'Pert PRO diffractometer equipped with Cu K$a$ radiation, using powdered X-ray diffraction. The FTIR spectra of pristine MgO and MgO/GNPs nanocomposites were recorded by Shimadzu (IR Tracer-100) spectrometer using KBr pellets method. The morphology and microstructure of the prepared samples was studied by field emission scanning electron microscopy (FE-SEM) (MIRA3 TESCAN) and TEM (JEOL 2100 F) operating at 200 kV. The Raman spectroscopic measurements of powdered samples were taken at room temperature using Ramboss equipped with the excitation laser of wavelength 314 nm. The photoluminescence measurements were made at room temperature with a laser of 325 nm wavelength. The UV-Vis spectra were recorded by UV-Vis spectrophotometer (Shimadzu Pharmaspec-1700).

## III. RESULTS AND DISCUSSION

### A. Structural and morphological analysis

The crystalline nature and phase purity of pristine MgO and MgO/GNPs nanocomposites was analysed in the range of 10 –80 in the diffractograms depicted in Fig. 1. The sample MgO shows peaks located at 36.9 (111), 42.7 (200), 61.9 (220), 74.5 (311), and 78.5 (222). All these peaks can be indexed to single phase cubic crystalline MgO structure and perfectly match with JCPDS-00-043-1021, with lattice parameters a = b = c = 4.2130 Å, and $\alpha = \beta = \gamma = 90º$. No peaks related to $Mg(OH)_2$ phase are observed. The X-ray diffractograms of MgO/GNPs with two different concentrations of GNPs show all the peaks of cubic MgO phase along with the small diffraction peak at 26º. This peak is the manifestation of C (002) plane contributed by the graphitic matrix. Thus, X-ray diffractograms confirm the formation of multiphase MgO/GNPs nanocomposites. The X-ray diffrac- tograms agree with previous reports.[39,40]

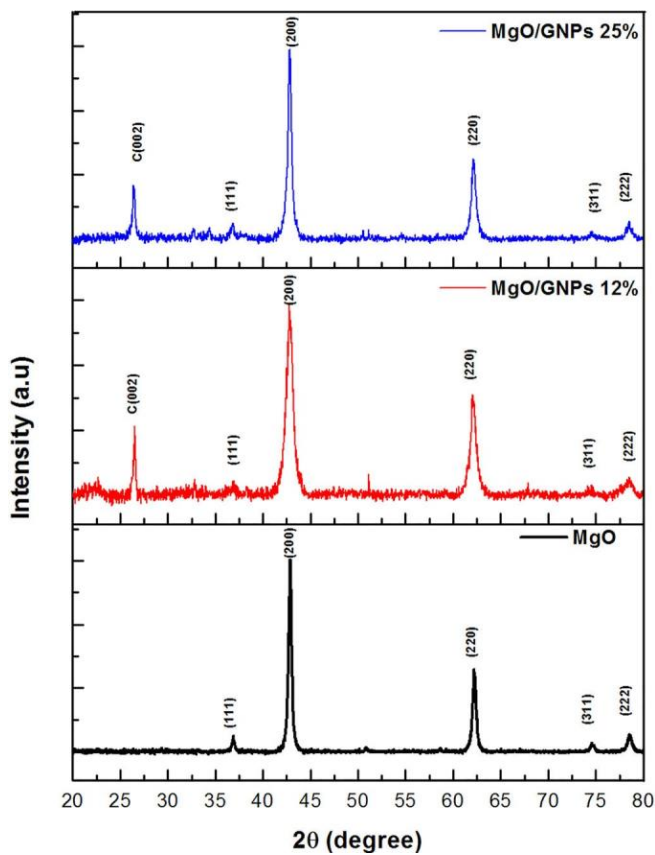

FIG. 1. X-ray diffractograms describing the crystalline phase of MgO and MgO/GNPs nanocomposites.

To investigate the morphology of MgO/GNPs nanocomposites, FE-SEM and TEM analysis was conducted. The images reveal the multiphase nature of material by formation of MgO nanohexagons on GNPs. Fig. 2(a) establishes that MgO is mainly composed of MgO units which have the morphology of lamellar to hexagons with variable edge lengths as labelled in Fig. 2(a). Figs. 2(b) and 2(c) represent the microstructure of MgO/GNPs nanocomposite with 25% GNPs loading. MgO nanohexagons seem to be embedded on GNPs. MgO nanohexagons with the edge lengths in the range of 121 nm–190 nm can be observed on graphene nanoplatelets. This assembly confirms the multiphase nature of the nanocomposites material. It is interesting to note that, on the inclusion of graphene nanoplatelets, the edge length of MgO nanohexagons reduces significantly as compared to the pristine MgO. The insets in Fig. 2 present the energy dispersive X-ray (EDX) spectra of MgO and MgO/GNPs 25% nanocomposites and confirm the elemental composition of prepared samples. The TEM image shown in Fig. 2(d) further verifies the multiphase nature of MgO/GNPs 25% nanocomposite. The HRTEM images are presented in Figs. 2(e)–2(g), whereas the insets are the selected area electron diffraction patterns. These images show the crystalline nature of MgO and agree with the results obtained from X-ray diffraction patterns. The HR-TEM images presented in Figs. 2(e) and 2(g) clearly show the interface between graphene sheets and MgO. It confirms the successful coupling between MgO and graphene sheets. To demonstrate the elemental confirmation of MgO/GNPs 25% nanocomposite, Fig. 2(h) presents the EDS spectrum. The presence of Mg, O, and C confirms the formation of the MgO/GNPs nanocomposite.

### B. FTIR and Raman analysis

To verify the chemical bond formation in the pristine MgO and nanocomposites, the FTIR spectroscopy was used to obtain the data in the range 400 cm$^{-1}$–4000 cm$^{-1}$. The FTIR spectroscopic response is elaborated in Fig. 3. MgO nanohexagons possess bands below 1000 cm$^{-1}$. In the case of pristine MgO nanohexagons, four bands are observed. A band at 424 cm$^{-1}$ can be attributed to the Mg-O stretching vibrations. The other bands associated with Mg-O vibrations are 538 cm$^{-1}$, 683 cm$^{-1}$, and 881 cm$^{-1}$. The adsorbed water molecules and surface hydroxyl groups manifest their presence by a band around 1442 cm$^{-1}$ and 1630 cm$^{-1}$. These strong bands appear due to the hygroscopic nature of MgO. The chemical bond Mg-OH has established the vibrations by a band around 3470 cm$^{-1}$. The MgO/GNPs nanocomposites with 12% and 25% loaded GNPs show a slight shift in the 538 cm$^{-1}$ band to higher wavenumbers. This shift is due to the incorporation of GNPs matrix in the nanocomposite material. Thus, FTIR results are in agreement with the previous report.[41]

Raman analysis is of utmost importance for the analysis of graphene based systems as it gives information about the quality of graphene and the direct evidence of formation of its nanocomposites with the other species. The Raman spectra of MgO/GNPs nanocomposites are presented in the range of 1200 cm$^{-1}$–1700 cm$^{-1}$ as shown in Fig. 4. The characteristic bands of graphene nanoplatelets are first order scattering bands, i.e., defect band (D-band) and G-band. The D-band is induced by the disorder present on $sp^2$ hybridized planar structure of graphene. It is primarily activated in the presence of lattice defects, doping, and covalent attachments in the form of functionalizations or other atoms.[42] The doubly degenerate $E_{2g}$ mode (G-band) appears due to first order scattering at 1587 cm$^{-1}$. The D-band of GNPs is located at 1350 cm$^{-1}$. In MgO/GNPs, 12% and 25% nanocomposites, the G-band exhibits shift to higher values by 19 cm$^{-1}$, i.e., to 1606 cm$^{-1}$. This shift is associated with charge transfer between MgO and GNPs, and depicts the formation of nanocomposites.[43] The D-band in MgO/GNPs 12% and 25% have become broader and an increase in its intensity is observed as compared to bare GNPs. This is due to the incorporation of MgO on the surface of GNPs. The intensity ratio between D-band and G-band, i.e., $I_D/I_G$ for GNPs is 0.73. It is increased to 0.941, and 0.953 for MgO/GNPs 12% and MgO/GNPs 25% nanocomposites, respectively. On the bases of Raman analysis, the incorporation of MgO with GNPs in the form of nanocomposites material can be speculated. The G-band shift and the increased $I_D/I_G$ values indicate the formation of nanocomposites.

### C. Photocatalytic properties

The photocatalytic activity of the synthesized samples, i.e., MgO and MgO/GNPs nanocomposites is investigated by using an industrial dye methyl orange as the water

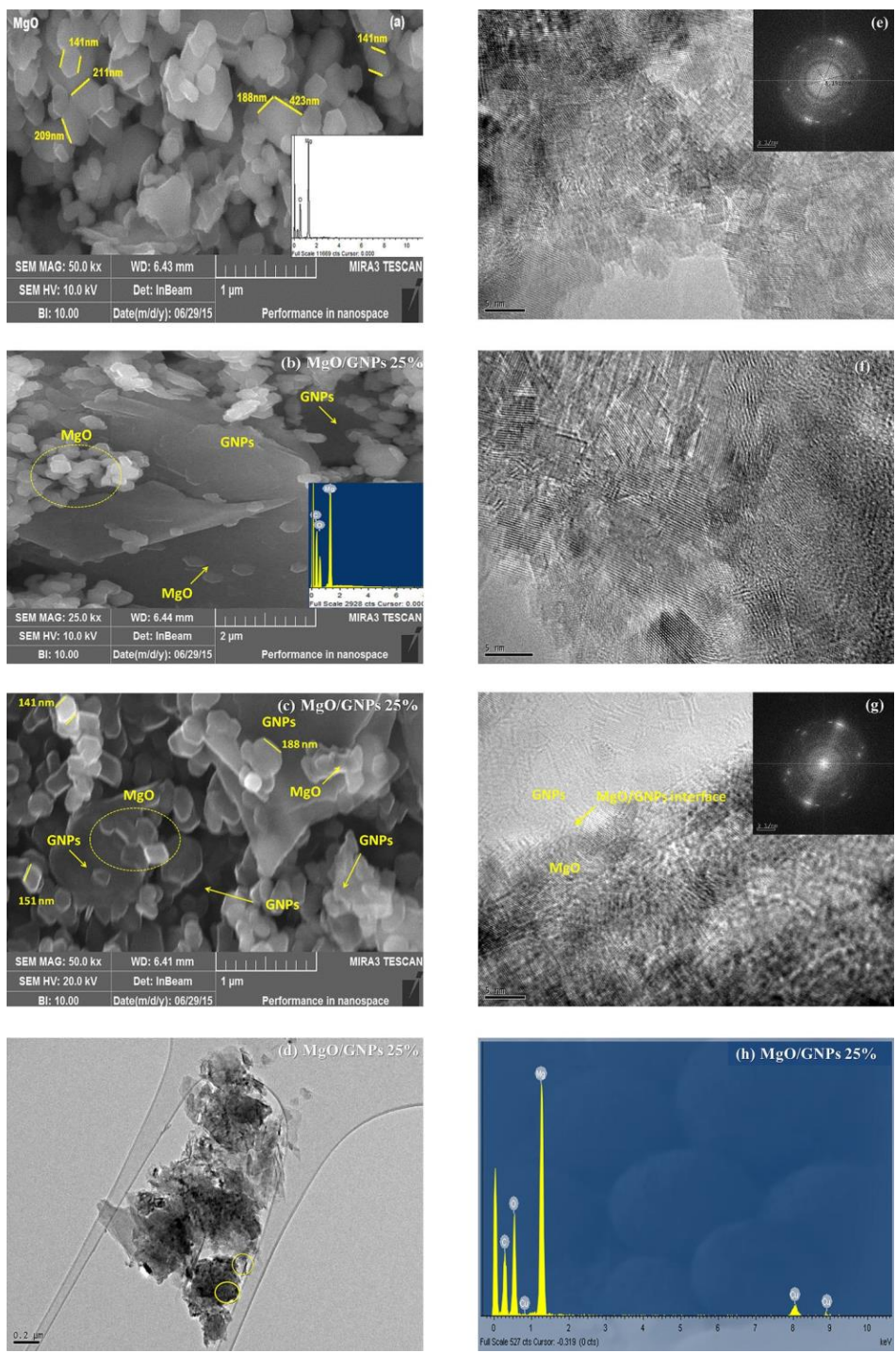

FIG. 2. SEM micrographs of (a) MgO (b), (c) MgO/GNPs 25% nanocomposites. Insets are the EDX spectra. (d) TEM image of MgO/GNPs 25%, (e)–(g) HR-TEM images of MgO and MgO/GNPs 25%, insets depict SAED patterns, and (h) EDS of MgO/GNPs 25% nanocomposite.

contaminant. All the samples were stirred in dark for 120 min and the adsorption-desorption equilibrium was achieved in 30 m as presented in Figs. 5(a)–5(c). The degradation of methyl orange strongly depends on the power of UV lamp used. The dye concentration used for the degradation process is optimized for the given power of UV lamp. If a higher concentration of the dye is used, then the time required for the degradation would be higher. Same is true if a low power light source is used. The UV-Visible absorption spectra for methyl orange solution under UV light irradiation in the presence of photocatalysts are shown in Figs. 6(a)–6(c). The absorbance of methyl orange showed a decreasing trend with an increase in irradiation time. The photocatalysts MgO and MgO/GNPs 12% nanocomposites photodegrade 43% and 44% dye, respectively, in 120 m. Methyl orange is degraded to 97.23% in the presence of MgO/GNPs 25% nanocomposite in the same time. The % degradation efficiencies are calculated using the expression,

% degradation efficiency = $(1 - C_t / C_o) \times 100$;

where $C_o$ is the initial concentration of aqueous solution of methyl orange, $C_t$ is its concentration at different time intervals, and $C_t$ and $C_o$ are determined by using the Beer-Lambert's law. The photocatalytic degradation of methyl orange in the presence of MgO and MgO/GNPs nanocomposites under UV light irradiation is shown in Fig. 7. The mechanism for the degradation of methyl orange is proposed in Fig. 8. The molecules of methyl orange can be transferred to the surface of the photocatalysts (i.e., by the adsorption process). After the UV light illumination, the valence electrons of MgO are excited to its conduction band. As the graphene family materials possess an excellent electrical conductivity, therefore these electrons are ultimately accepted by adjacent GNPs network. Parallel to this, an increase in the number of holes is observed. As GNPs are the good acceptors of electrons, so the conductive network of GNPs retains the charge carriers, ultimately delaying the recombination of $e_{CB}^{þ}$-$h_{VB}^{þ}$ pairs. The increase in the number of holes and electrons initiate the generation of reactive oxygen species intermediates are then converted into $CO_2$, and $H_2O$ as explained step by step in Fig. 9. It leads to the complete degradation of methyl orange. The suggested mechanism depicting the formation of intermediates is already proven with evidences in previous studies.[47,48] The photocatalytic reaction can be explained by following reactions:

$$MgO - GNPs + h\nu \rightarrow MgO\,(h^+{}_{VB}) - GNPs\,(e^-{}_{CB})$$

$$MgO\,(h^+{}_{VB}) + (H_2O \rightarrow H^+ + OH^-) \rightarrow MgO + H^+ + \bullet OH$$

$$GNPs\,(e^-{}_{CB}) + O_2 \rightarrow GNPs + O^-{}_2$$

$$O^-{}_2 + (H^+ + OH^-) \rightarrow H_2O\bullet + OH^-$$

$$ROS + Methyl\ orange \rightarrow H_2O + CO_2$$

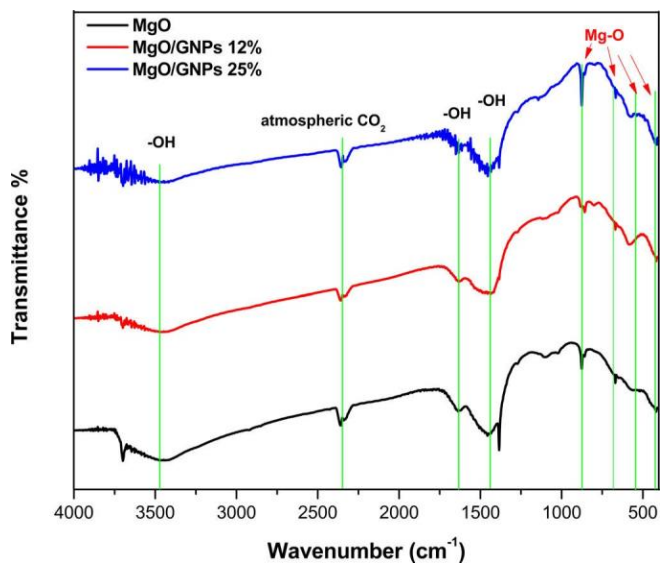

FIG. 3. FTIR spectra of MgO/GNPs nanocomposites.

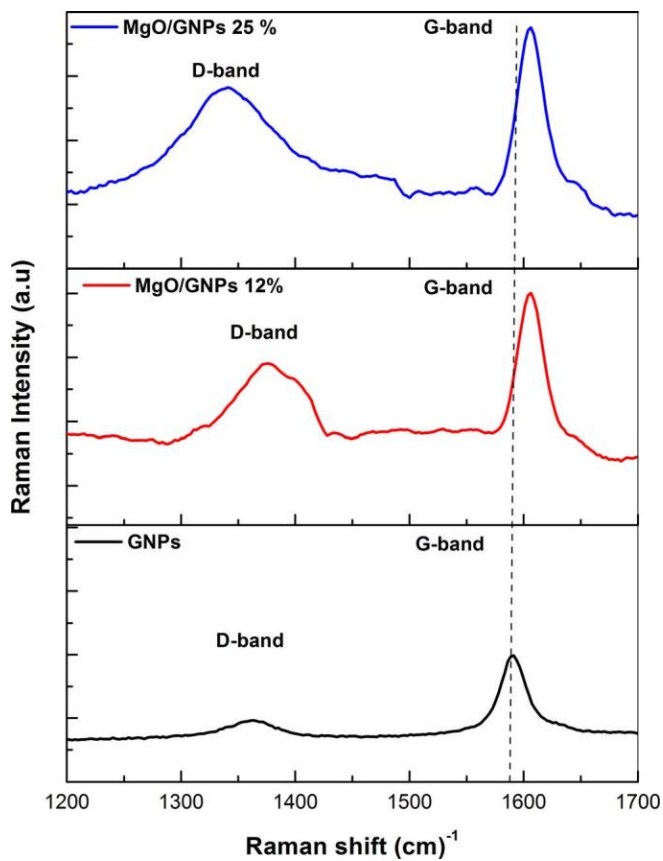

FIG. 4. Raman analysis of GNPs, MgO/GNPs 12%, and MgO/GNPs 25% nanocomposites.

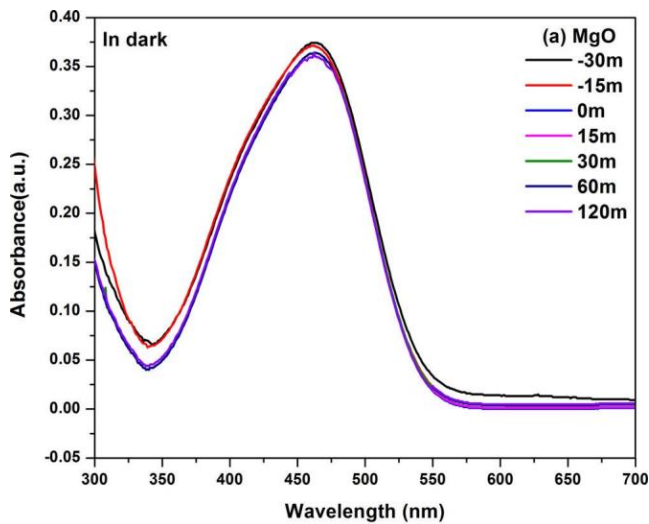

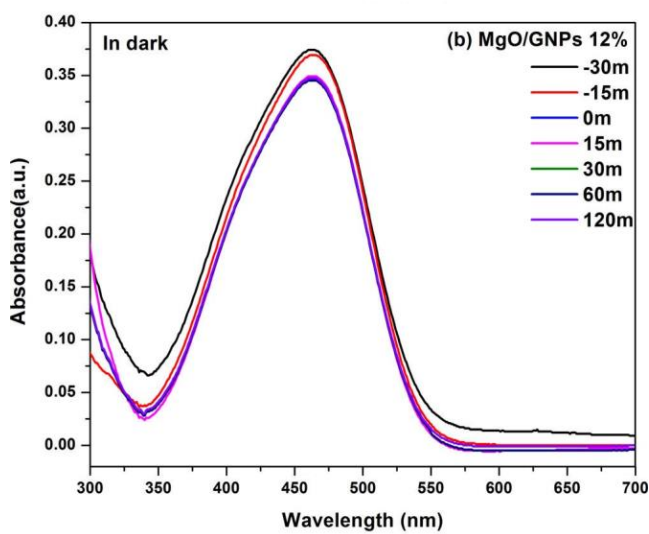

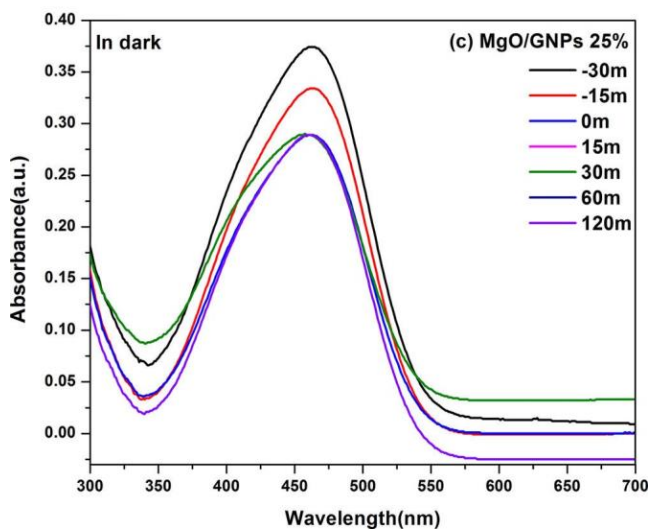

FIG. 5. (a)–(c) Adsorption-desorption equilibrium curves of MgO, MgO/GNPs 12%, and MgO/GNPs 25% nanocomposites.

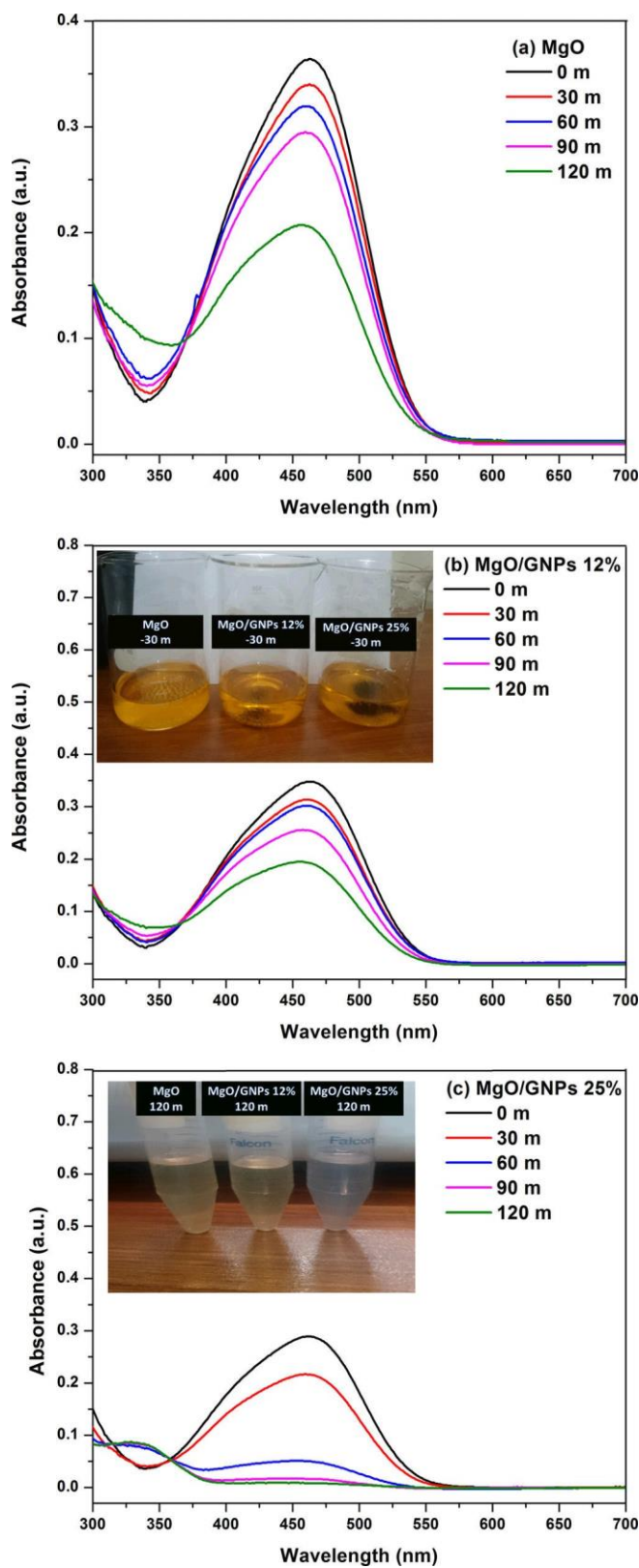

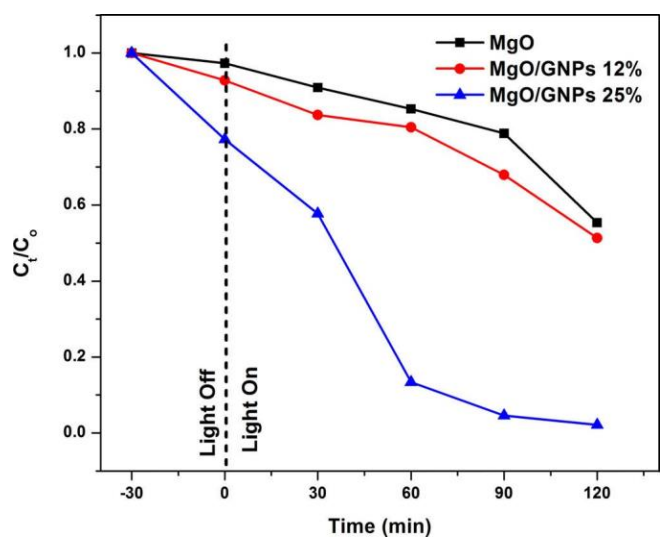

FIG. 7. Photodegradation curves of MgO and MgO/GNPs nanocomposites.

The MgO/GNPs nanocomposites show a better photocatalytic activity than bare MgO. The importance of the optimum percentage of GNPs in nanocomposite is exhibited by the excellent enhancement in photocatalytic activity of MgO/GNPs 25% nanocomposite. From these observations, the highest photocatalytic activity associated with maximum GNPs content is attributed to the effective electron transfer from the conduction band of MgO to GNPs. This transfer

FIG. 6. (a)–(c) Time evolution of absorbance spectra of methyl orange in the presence of various photocatalysts.

(ROS). It is well established in the previous literature that the holes produced in the valence band react with chemisorbed water to produce OH radicals.[44–46] These radicals successively attack methyl orange. It results in the oxidation of adsorbed dye by producing various intermediates. The

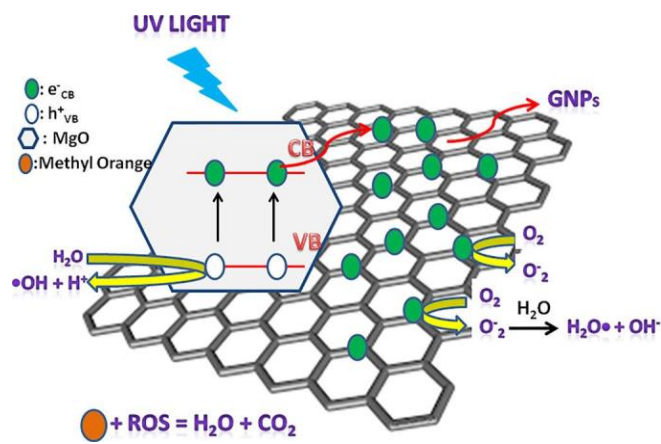

FIG. 8. Schematic presentation of the photocatalytic activity of MgO/GNPs nanocomposites.

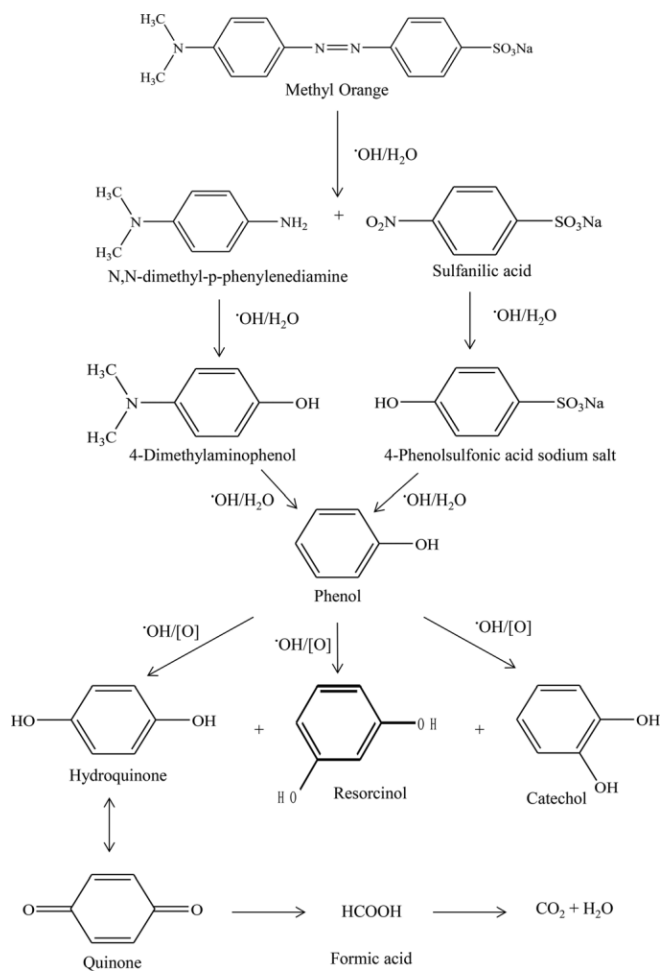

FIG. 9. Schematic illustration of the conversion of methyl orange to intermediates in the presence of reactive oxygen species.

reduces the recombination of photogenerated $e^-_{CB}$-$h^+_{VB}$ pairs. The steady state photoluminescence analysis was conducted for MgO and MgO/GNPs 25% nanocomposite to examine the charge carrier trapping and recombination process. The generation and separation of the charge carriers is the key factor that influences the photocatalytic response of MgO and MgO/GNPs nanocomposite. The photoluminescence intensity indicates the recombination of charge carriers. Fig. 10 depicts that MgO possess a much stronger intensity which is an indication of fast recombination of photogenerated $e^-_{CB}$-$h^+_{VB}$ pairs. Whereas the inclusion of graphene nanoplatelets significantly reduces the PL intensity. The charge trapping induced by GNPs provides clue for an excellent photocatalytic activity of MgO/GNPs 25% nanocomposite. This behavior is reported previously as well.[49,50] Moreover the remarkable increase of photocatalytic activity of MgO/GNPs nanocomposite can also be attributed to the strong interaction between MgO and defect sites of GNPs and MgO nanocomposite. In order to thoroughly investigate the photocatalysis process, the apparent rate constants of the reactions are determined by applying pseudo first order kinetics. The rate constants are determined by expression, $\ln(C_o/C_t) = kt$

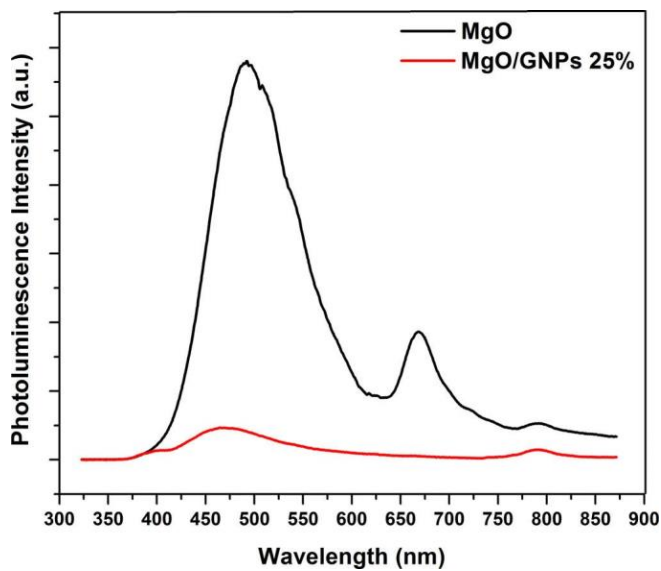

FIG. 10. Photoluminescence spectra of MgO and MgO/GNPs 25% nanocomposite.

Here k is the apparent rate constant, and is obtained by linear fitting of the data as presented in Fig. 11. The apparent rate constant increases significantly for MgO/GNPs 25% (0.02 m$^{-1}$) where as it is 0.003 m$^{-1}$ and 0.004 m$^{-1}$ for MgO and MgO/GNPs 12%, respectively. The photocatalytic findings suggest a way for the fast and efficient degradation of methyl orange by modification of MgO. These results are extremely better than the previous study where methyl orange was degraded up to 92% and 96% in 270 m and 210 m, respectively.[51]

It is essentially important to study the stability and durability of the photocatalysts for practical benefits. The recyclability performance of MgO/GNPs 25% nanocomposite was studied under similar conditions. The inset in Fig. 11 shows that the MgO/GNPs 25% nanocomposite shows no significant loss of photocatalytic activity during three successive UV activated degradation experiments. The results indicate that MgO/GNPs 25% nanocomposite is an excellent photocatalyst under the UV light irradiation for practical benefits. Here results clearly show that methyl orange could be removed up to 97.23% by the photocatalyst MgO/GNPs 25% nanocomposite in much less time, i.e., 120 m with a high apparent rate constant. It is also found that the efficient photodegradation of methyl orange is achieved with a less amount of photocatalyst (0.4 g/l) in the present study where as 0.5 g/l (0.1 g/200 ml) MgO was used to decolorize the methyl orange in a previous work.[52]

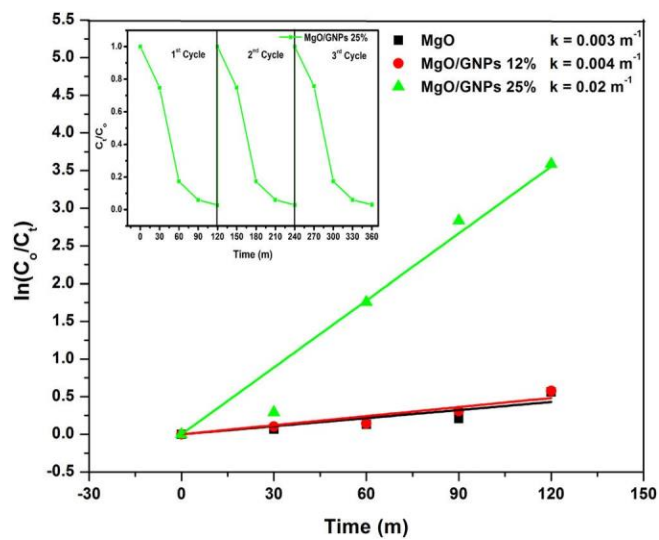

FIG. 11. Pseudo first order kinetics of degradation of methyl orange in the presence of MgO and MgO/GNPs nanocomposites. Inset is the recyclability performance of MgO/GNPs 25% nanocomposite.

## D. Antibacterial properties

MgO is considered as a nontoxic material for the human and animal tissue as it is used as anti-laxative and relieving agent of stomach burn. It has a well known antibacterial activity.[13,20–22] In this study, the effect of increased amount of GNPs on the antibacterial activity of MgO/GNPs nanocomposites is comprehensively investigated. The antibacterial activity of MgO/GNPs nanocomposites is evaluated against both Gram positive (Methicillin resistant *Staphylococcus aureus*) and Gram negative bacterial strains (*Pseudomonas aeruginosa,* and *Escherichia coli*). The results are obtained by analysing the bacterial strains in the absence and presence of aqueous colloidal suspensions of MgO/GNPs nanocomposites for an incubation time of 24 h. Absorbance at 600 nm was recorded up to 24 h to monitor the growth profile of the bacteria in the presence and absence of the test samples. The results of time kill assay are presented by growth inhibition curves in Fig. 12. The control sample represents the untreated bacterial strains. The nanocomposites show an antibacterial activity towards all the three bacterial strains. It is found that the growth of *S. aureus* is significantly inhibited in the presence of MgO/GNPs 25% nanocomposite as compared to MgO/GNPs 12%, presented as histograms in Fig. 12 (49% and 25%, respectively). For *E. coli*, the MgO/GNPs 12% and MgO/GNPs 25% nanocomposites inhibit 33% and 44.5% growth rate, respectively. The findings further reveal that 22% and 22.38% growth of *Pseudomonas aeruginosa* is inhibited by MgO/GNPs 12% and MgO/GNPs 25% nanocomposites, respectively. Thus, the time kill assay suggests that MgO/GNPs 25% nanocomposite has been found more effective as compared to MgO/GNPs 12% nanocomposite in controlling the bacterial growth.

The increased amount of GNPs in the nanocomposites results in a higher inhibition of growth. MgO/GNPs 25% has the highest inhibition growth rate for *S. aureus*, i.e., 49% inhibition of bacterial growth is achieved. So, it is slightly more effective to inhibit the growth rates for Gram positive bacterial strains. A comparative analysis is presented in Table I, that catalogues the results from other studies that explore the antibacterial performance of the graphene based materials.

The exact mechanism responsible for the loss of bacterial integrity is still under debate. In previous studies, various mechanisms have been proposed to account for reduction in the growth rate of pathogenic bacteria. The MgO decorated sheets like structure of GNPs (as confirmed by FE-SEM and TEM images in Figs. 2(b) and 2(d)) demonstrates that bacterial strains may develop an intimate contact with the rough surface of MgO loaded GNPs. The edges of GNPs exert stress on the bacterial cell wall, leading to rupturing of the cell membranes with an ultimate leakage of the bacterial cytoplasmic content. These edges play the role of cutters for cell membranes. It results in the bacterial cell death.[53,59] Similar antibacterial mechanism involving physical contact of tubes with bacteria and internalization of small tubes has been suggested for single wall carbon nanotubes, multiwall carbon nanotubes, and fullerenes as well. All these species are chemically like graphene.[60–63] It may be further suggested that wrapping of sheet like nanocomposites around bacteria provides a higher concentration of metal oxide nanostructures (MgO in our case) on the bacterial surface. It is elucidated by Raman spectra in Fig. 4, that nanocomposites are rich in defects. The rough surface of MgO decorated GNPs envelops the bacteria. The loss of bacterial membrane integrity is the consequence of accumulations of MgO nanohexagons.[64]

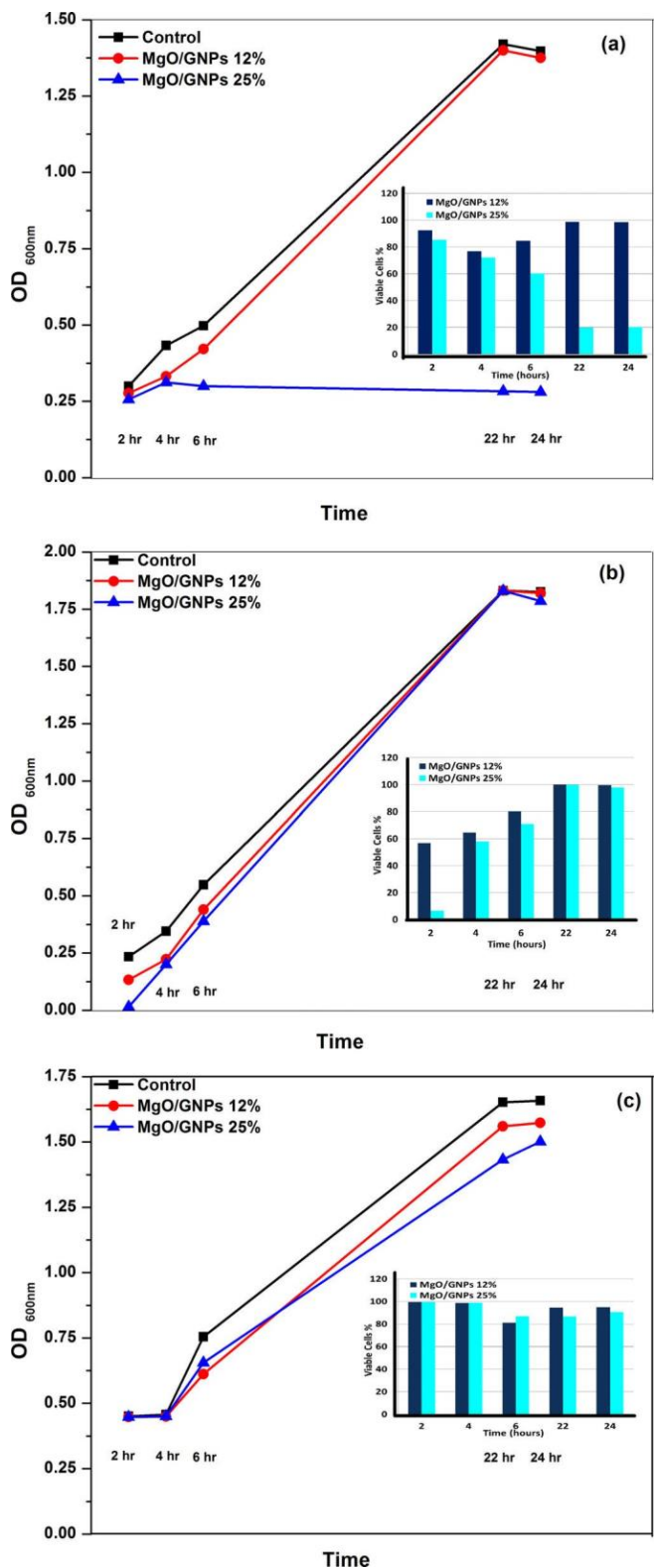

FIG. 12. Effect of MgO/GNPs nanocomposites on the growth of different bacterial strains (a) *S. aureus*, (b) *E. Coli*, and (c) *Pseudomonas aeruginosa*. Insets depict the cell viability analysis in the presence of MgO/GNPs nanocomposites.

TABLE I. A comparative summary of antibacterial performance of graphene based nanomaterials.

| Material | % Cell inactivation S. aureus | % Cell inactivation E. coli | % Cell inactivation Pseudomonas aeruginosa | References |
|---|---|---|---|---|
| Graphite | Not tested | 26.1 ± 4.8% | Not tested | 53 |
| Graphite oxide | Not tested | 15.0 ± 3.7% | Not tested | 53 |
| Graphene oxide | Not tested | 69.3 ± 6.1% | Not tested | 53 |
| rGO | Not tested | 49.5 ± 4.8% | Not tested | 53 |
| Graphene oxide film | 61% | 51% | Not tested | 54 |
| Polyethyleneimine-modified rGO | 20.5 ± 0.9% | 14.8 ± 1.7% | Not tested | 55 |
| Graphene on Cu | 34% | 56% | Not tested | 56 |
| Graphene oxide | Not tested | Growth is enhanced | Not tested | 57 |
| Graphite | <10% | Not tested | <10% | 58 |
| Graphene-R | 49% | Not tested | 83% | 58 |
| MgO/GNPs 25% | 49% | 44.5% | 22.4% | Present study |

Another plausible mechanism is suggested recently for graphene induced cell death. In few recent researches, it has been additionally reasoned that the charge imbalance on cell membrane leads to membrane collapse. As the cell membranes of bacteria are negatively charged. Graphene family nanostructures are considered as good electron acceptors due to their exceptionally high electrical conductivity. The contact between bacteria and nanocomposite may result in the flow of negative charge from bacterial membrane to the GNPs which provide a conductive network enveloping the cell membranes. This leads to charge imbalance on the bacterial cell membrane, thus inducing the bacterial death.[65] Some studies on other metal oxide/graphene based nanocomposites have suggested that the release of positive metal ions ($Mg^+$ ions from MgO nanohexagons in present study) followed by penetration of ions into the cell membranes eventually induces bacterial death.[66]

In the light of above discussion, it is suggested that superior antibacterial performance of MgO/GNPs nanocomposites is the synergistic effect of edge stress created by GNPs, $Mg^+$ ions internalization by bacteria and charge imbalance created on bacterial cell boundary (due to flow of negative charge from the membrane to GNPs). Gram negative bacte- ria have been found to be relatively more resistant as com- pared to the Gram positive, S. aureus bacteria.

S. aureus is antibiotic resistant and is potentially harmful. It spreads very easily on a direct contact with the infected person or contaminated objects. E. coli are generally found in intestine and can be transferred from unsafe drinking water and food. They may cause intestinal infections.[38] Therefore, it is essential to develop a cheap, easy to produce, and effective antibacterial agents giving control over growth of pathogenic bacteria. In this regard, MgO/GNPs 25% nanocomposite is an effective antibacterial agent. We envision that the antibacterial performance of MgO/GNPs nanocomposites can further be tailored by the variation of dose rates during incubation and by changing the amount of GNPs in the nanocomposites.

## IV. CONCLUSIONS

The MgO/GNPs nanocomposites with different concentrations of GNPs have been synthesized successfully by a simple solvothermal route. The electron accepting nature of GNPs plays the vital role for making MgO/GNPs 25% nanocomposite bifunctional. The assembly of GNPs and MgO provides an efficient route to enhance the photocatalytic properties of MgO nanohexagons up to 97.23% in 120 min under UV light irradiation. The investigation of GNPs amount dependent antibacterial activity reveals that MgO/GNPs 25% nanocomposite shows a higher toxicity towards S. aureus and E. coli with 49% and 44.5% inhibition of growth, respectively. Therefore, the prepared nanocomposite MgO/GNPs 25% can be used as a multifunctional material for cleaning of waste water and antibacterial agent.

## ACKNOWLEDGMENTS

We are thankful for the valuable suggestions of the anonymous reviewers, for the improvement of this manuscript. We acknowledge the support provided by Dr. Ian Terry, Dr. G. B. Mendis, Dr. Douglas Halliday, and Dr. Mahavir Sharma, Department of Physics, Durham University, UK. The research is funded by the Higher Education Commission of Pakistan (HEC) IRSIP (Grant No.: 1-8/HEC/HRD/2016/5995 PIN: IRSIP 32 PSc 04) and the Higher Education Commission of Pakistan (HEC) NRPU (Grant No.: 20-4861/R&D/HEC/14) to Aqsa Arshad and Dr. Javed Iqbal, respectively.

[1] A. Sarkar, E. G. Espino, T. Wågberg, A. Shchukarev, M. Mohl, A. R. Rautio, O. Pitkänen, T. Sharifi, K. Kordas, and J. P. Mikkola, Nano Res. 9, 1956–1968 (2016).
[2] L. Pan, S. Wang, J. Xie, L. Wang, X. Zhang, and J. J. Zou, Nano Energy 28, 296–303 (2016).
[3] S. Bai, L. Wang, X. Chen, J. Du, and Y. Xiong, Nano Res. 8, 175–183 (2015).
[4] W. J. Ong, L. L. Tan, S. P. Chai, S. T. Yong, and A. R. Mohamed, ChemSusChem 7, 690–719 (2014).
[5] S. Bai, M. Xie, Q. Kong, W. Jiang, R. Qiao, Z. Li, J. Jiang, and Y. Xiong, Part. Part. Syst. Charact. 33, 506–511 (2016).
[6] D. Spagnoli, J. P. Allen, and S. C. Parker, Langmuir 27, 1821–1829 (2011).
[7] D. M. Roessler and W. C. Walker, Phys. Rev. 159, 733–738 (1967).
[8] C. Y. Tai, C. T. Tai, M. H. Chang, and H. S. Liu, Ind. Eng. Chem. Res. 46, 5536–5541 (2007).
[9] X. S. Fang, C. H. Ye, L.-D. Zhang, J. X. Zhang, J. W. Zhao, and P. Yan, Small 1, 422–428 (2005).
[10] Y. Yin, G. Zhang, and Y. Xia, Adv. Funct. Mater. 12, 293–298 (2002).
[11] Y. B. Li, Y. Bando, D. Golberg, and Z. W. Liu, Appl. Phys. Lett. 83, 999–1001 (2003).
[12] Z. X. Tang and B. F. Lv, Braz. J. Chem. Eng. 31, 591–601 (2014).
[13] F. Al-Hazmi, F. Alnowaiser, A. A. Al-Ghamdi, A. A. Al-Ghamdi, M. M. Aly, R. M. Al-Tuwirqi, and F. El-Tantawy, Superlattices Microstruct. 52, 200–209 (2012).
[14] B. Vatsha, P. Tetyana, P. M. Shumbula, J. C. Ngila, L. M. Sikhwivhilu, and R. M. Moutloali, J. Biomater. Nanobiotechnol. 4, 365–373 (2013).
[15] J. Lee, T. Jeong, S. Yu, S. Jin, J. Heo, W. Yi, and J. M. Kim, J. Vac. Sci. Technol., B 19, 1366–1369 (2001).
[16] B. Q. Wei, R. Vajtai, Z. J. Zhang, G. Ramanath, and P. M. Ajaya, J. Nanosci. Nanotechnol. 1, 35–38 (2001); X. F. Fang, C. H. Ye, T. Xie, Z. Y. Wang, J. W. Zhao, and Li. Zhang, Appl. Phys. Lett. 88, 013101 (2006); R. Li, C. Xiong, D. Kaung, L. Dong, Y. Lei, J. Yao, M. Jaing, and L. Li, Macromol. Rapid Commun. 29, 1449–1454 (2008).
[17] L. Sohrabi, F. Taleshi, and R. Sohrabi, J. Mater. Sci.: Mater. Electron. 25, 4110–4114 (2014).
[18] T. G. Venkatesha, R. Viswanatha, Y. A. Nayaka, and B. K. Chethana, Chem. Eng. J. 198–199, 1–10 (2012).
[19] G. Moussavi and Mahmoudi, J. Hazard. Mater. 168, 806–812 (2009).
[20] J. Suresha, R. G. R. Gandhib, S. Selvamc, and M. Sundrarajan, Adv. Mater. Res. 678, 297–300 (2013).
[21] M. Sundrarajan, J. Suresh, and R. R. Gandhi, Dig. J. Nanomater. Biostruct. 3, 983–989 (2012).
[22] Y. H. Leung, A. M. C. Ng, X. Xu, Z. Shen, L. A. Gethings, M. T. Wong, C. M. N. Chan, M. Y. Guo, Y. H. Ng, A. B. Djurisić, P. K. H. Lee, W. K. Chan, L. H. Yu, D. L. Phillips, A. P. Y. Ma, and F. C. C. Leung, Small 10, 1171–1183 (2014).
[23] K. S. Novoselov, A. K. Geim, S. V. Morozov, D. Jiang, Y. Zhang, S. V. Dubonos, I. V. Grigorieva, and A. A. Firsov, Science 306, 666–669 (2004).
[24] A. H. C. Neto, F. Guinea, N. M. R. Peres, K. S. Novoselov, and A. K. Geim, Rev. Mod. Phys. 81, 109–162 (2009).
[25] C. N. R. Rao, A. K. Sood, K. S. Subrahmanyam, and A. Govindaraj, Angew. Chem., Int. Ed. 48, 7752–7777 (2009).
[26] A. K. Geim and K. S. Novoselov, Nat. Mater. 6, 183–191 (2007).
[27] R. Muszynski, B. Seger, and P. V. Kamat, J. Phys. Chem. C 112, 5263–5266 (2008).
[28] Y. Y. Wen, H. M. Ding, and Y. K. Shan, Nanoscale 3, 4411–4417 (2011).
[29] H. Seema, K. C. Kemp, V. Chandra, and K. S. Kim, Nanotechnology 23, 355705–355712 (2012).
[30] S. Stankovich, D. A. Dikin, G. H. B. Dommett, K. M. Kohlhaas, E. J. Zimney, E. A. Stach, R. Piner, S. T. Nguyen, and R. S. Ruoff, Nature 442, 282–286 (2006).
[31] T. S. Sreeprasad, S. M. Maliyekkal, K. P. Lisha, and T. Pradeep, J. Hazard. Mater. 186, 921–931 (2011).
[32] C. Xu, X. Wang, and J. Zhu, J. Phys. Chem. C 112, 19841–19845 (2008).
[33] W. J. Ong, L. L. Tan, S. P. Chai, and S. T. Yong, Dalton Trans. 44, 1249–1257 (2015).
[34] W. J. Ong, L. L. Tan, Y. H. Ng, S. T. Yong, and S. P. Chai, Chem. Rev. 116, 7159–7329 (2016).
[35] J. R. Lee and H. Y. Koo, Carbon Lett. 14, 206–209 (2013).
[36] J. R. Lee, J. Y. Bae, W. Jang, J. H. Lee, W. S. Choi, and H. Y. Koo, RSC Adv. 5, 83668–83673 (2015).
[37] F. P. Du, W. Yang, F. Zhang, C. Y. Tang, S. P. Liu, L. Yin, and W. C. Law, ACS Appl. Mater. Interfaces 7, 14397–14403 (2015).
[38] T. Jan, J. Iqbal, M. Ismail, and A. Mahmood, J. Appl. Phys. 115, 154308 (2014).
[39] H. W. Kim, S. H. Shim, J. W. Lee, and C. Lee, J. Korean Phys. Soc. 51, 204–208 (2007).
[40] H. R. Moon, J. J. Urban, and D. J. Milliron, Angew. Chem., Int. Ed. 48, 6278–6281 (2009).
[41] J. Han, B. K. Woo, W. Chen, M. Sang, X. Lu, and W. Zhang, J. Phys. Chem. C 112, 17512–17516 (2008).
[42] A. Gupta, G. Chen, P. Joshi, S. Tadigadapa, and P. C. Eklund, Nano Lett. 6, 2667–2673 (2006).
[43] M. S. A. S. Shah, A. R. Park, K. Zhang, J. H. Park, and P. J. Yoo, ACS Appl. Mater. Interfaces 4, 3893–3901 (2012).
[44] M. Y. Guo, A. M. C. Ng, F. Liu, A. B. Djurisic, and W. K. Chan, Appl. Catal., B 107, 150–157 (2011).
[45] Y. Badr and M. A. Mahmoud, J. Phys. Chem. Solids 68, 413–419 (2007).
[46] F. Soofivand and M. S. Niasari, RSC Adv. 5, 64346–64353 (2015).
[47] R. Comparelli, E. Fanizzaa, M. L. Curri, P. D. Cozzoli, G. Mascolo, R. Passino, and A. Agostiano, Appl. Catal., B 55, 81–91 (2005).
[48] R. Comparelli, E. Fanizza, M. L. Curri, P. D. Cozzoli, G. Mascolo, and A. Agostiano, Appl. Catal., B 60, 1–11 (2005).
[49] S. Gayathri, P. Jayabal, M. Kottaisamy, and V. Ramakrishnan, J. Appl. Phys. 115, 173504 (2014).
[50] X. An, J. C. Yu, J. Wang, Y. Hu, X. Yu, and G. Zhang, J. Mater. Chem. 22, 8525–8531 (2012).
[51] K. Mageshwari, S. S. Mali, R. Sathyamoorthy, and P. S. Patil, Powder Technol. 249, 456–462 (2013).
[52] S. F. Bdewi, A. M. Abdulrazaka, and B. K. Aziz, Asian Trans. Eng. 5, 1–5 (2015).
[53] S. Liu, T. H. Zeng, M. Hofmann, E. Burcombe, J. Wei, R. Jiang, J. Kong, and Y. Chen, ACS Nano 5, 6971–6980 (2011).
[54] Q. Bao, D. Zhang, and P. Qi, J. Colloid Interface Sci. 360, 463–470 (2011).
[55] X. Cai, M. Lin, S. Tan, W. Mai, Y. Zhang, Z. Liang, Z. Lin, and X. Zhang, Carbon 50, 3407–3415 (2012).
[56] L. Dellieu, E. Lawaree, N. Reckinger, C. Didembourg, J. J. Letesson, M. Sarrazin, O. Deparis, J. Y. Matrouleb, and J. F. Colomer, Carbon 84, 310–316 (2015).
[57] O. N. Ruiz, K. A. S. Fernando, B. Wang, N. A. Brown, P. G. Luo, N. D. McNamara, M. Vangsness, Y. P. Sun, and C. E. Bunker, ACS Nano 5, 8100–8107 (2011).
[58] V. T. H. Pham, V. K. Truong, M. D. J. Quinn, S. M. Notley, Y. Guo, V. A. Baulin, M. A. Kobaisi, R. J. Crawford, and E. P. Ivanova, ACS Nano 9, 8458–8467 (2015).
[59] O. Akhavan and E. Ghaderi, ACS Nano 4, 5731–5736 (2010).
[60] A. M. Jastrzebska, P. Kurtycz, and A. R. Olszyna, J. Nano Res. 14, 1320–1341 (2012).
[61] C. D. Vecitis, K. R. Zordow, S. Kang, and M. Elimelech, ACS Nano 4, 5471–5479 (2010).
[62] S. Kang, M. Herzberg, D. F. Rodrigues, and M. Elimelech, Langmuir 24, 6409–6413 (2008).
[63] S. B. Liu, S. Wel, L. Hao, N. Fang, W. M. Chang, R. Xu, H. Y. Yang, and Y. Chen, ACS Nano 3, 3891–3902 (2009).
[64] Y. D. Lyon and J. J. P. Alvarez, Environ. Sci. Technol. 42, 8127–8132 (2008).
[65] Y. W. Wang, A. Cao, Y. Jiang, X. Zhang, J. H. Liu, Y. Liu, and H. Wang, ACS Appl. Mater. Interfaces 6, 2791–2798 (2014).
[66] J. Li, G. Wang, H. Zhu, M. Zhang, X. Zheng, Z. Di, X. Liu, and X. Wang, Sci. Rep. 4, 4359 (2014).